\documentclass{elsart1p}
\usepackage{graphicx}
\usepackage{amssymb}
\usepackage{subfigure}
\usepackage{threeparttable} 

\newcommand \sqn{$\sqrt{s_{_{NN}}}$ }
\begin{document}
\begin{frontmatter}
\title{Low Mass Vector Meson Measurements via Di-electrons at RHIC by
  the PHENIX Experiment}
\author{Deepali Sharma for the PHENIX Collaboration}
%\ead{deepali.sharma@weizmann.ac.il}
\address{Weizmann Institute of Science, Rehovot, 76100, Israel }
\begin{abstract}
  The PHENIX experiment at RHIC has measured $\omega$ and $\phi$
  mesons in $p+p$, $d+Au$ and $Au+Au$ collisions at \sqn = 200
  GeV via both hadronic and di-electron decay channels. The transverse
  momentum spectra as measured in different decay modes and at
  different centralities are shown and discussed here.
\end{abstract}
\end{frontmatter}
%\vspace{-0.2cm}
%\section{Introduction}
%\vspace{-0.3cm}
%\label{intro}

Low Mass Vector mesons are considered to be one of the most
interesting probes to study the properties of strongly interacting 
matter created in heavy-ion collisions. A precise knowledge of
their production rates and spectral properties are important in
understanding the medium properties. Due to their short lifetimes,
\textit{e.g.}, 23 $fm/c$ for $\omega$ and 46 $fm/c$ for $\phi$, a
considerable fraction of them decays inside the fireball, thus
providing information about in-medium modifications of their spectral
shape that could be linked to the restoration of chiral
symmetry. These changes can be observed directly without any
distortion through their dilepton decay modes as dileptons interact
only electromagnetically and have a relatively large mean free
path compared to the size of the system. These modifications could
manifest themselves as changes in the spectral shapes or in the
branching ratios. In particular, since $m_{\phi} \approx 2m_{K}$, even
small changes in the spectral properties of $\phi$ or $K$ can induce 
significant changes in the $\phi \rightarrow K^+K^-$ yield.
% \vspace{-0.3cm}
% \section{Experimental Set-up and Data Analysis}
% \label{analysis}
% \vspace{-0.4cm}

The PHENIX detector~\cite{Phenix} at RHIC is a versatile detector that
has the potential to measure the LVM properties. The analyses
presented here were performed using the two central arms of the PHENIX
spectrometer, each of them covering a pseudorapidity range $|\eta| ~<$
0.35 and 90$^{\circ}$ in azimuthal angle. The momentum of the charged
tracks is measured with high resolution Drift Chambers (DC) and the
first layer of Pad Chambers (PC1). Valid DC-PC1 tracks are confirmed by
the matching between the projected and associated hit information to
the Ring Imaging $\check{C}$erenkov (RICH) detector and Electromagnetic
Calorimeter (EMCal) in the case of electrons and to the Time of Flight
(TOF) detector or EMCal for the hadrons. Beam Beam Counters (BBC) and
Zero Degree Calorimeters provide the minimum bias trigger, the vertex
position and the centrality of the collision.  

The $\phi,~\omega \rightarrow e^+e^-$ and $\phi \rightarrow K^+K^-$
decays were reconstructed by combining the oppositely charged
identified particles into pairs to form invariant mass spectra that
contain both the signal and a combinatorial background of uncorrelated
pairs. The shape and amount of the combinatorial background in $Au+Au$
($\phi \rightarrow e^+e^-, K^+K^-$) and $d+Au$ ($\phi \rightarrow
e^+e^-$) analyses, were estimated using an event mixing procedure. The
spectra, after subtracting the combinatorial background, were then fit
in the vicinity of the meson masses to a Breit-Wigner (BW) function
convoluted with a Gaussian to take into account the detector
resolution, to extract the raw yield. For other analyses, the yield extraction was done by fitting
the raw peaks to a BW function convoluted with a Gaussian for the signal
and a polynomial of order two for the underlying background. The $\phi \rightarrow
K^+K^-$ analysis in $p+p$ and $d+Au$ was done using three different
techniques. One required both charged tracks to be identified as kaons
using the TOF or EMCal detectors, the second  one required only one to
be identified as kaon in the TOF and the last one assumes that all the
charged tracks are kaons. Results of all the three techniques show good
agreement to each other as can be seen in Fig. 1(a). The
$\omega \rightarrow \pi^{0}\gamma$ and $\omega \rightarrow
\pi^{0}\pi^+\pi^-$ decays were reconstructed by combining $\pi^0$
(reconstructed via $\pi^0 \rightarrow \gamma\gamma$) candidates with
all other photons from the same event or with a pair of oppositely
charged unidentified tracks, respectively. The raw yields were
extracted by fitting the invariant mass distributions around the $\omega$
peak with a Gaussian for the signal and a parabolic function for
the background. The extracted raw yields are corrected for reconstruction
efficiency that includes geometrical acceptance, track reconstruction,
detector performance and analysis cuts. The correction function is
determined using an event generator and a GEANT based simulation of
the PHENIX detector.

Fig. 1(a) shows the $\phi$ invariant $m_{T}$ spectra measured 
via $K^+K^-$ and $e^+e^-$ decay modes in $p+p$ and $d+Au$
collisions. The two decay modes are in reasonable agreement to each 
other in both systems. The limited statistics in $\phi \rightarrow
e^+e^-$ in $d+Au$ do not allow us to go to high $p_{T}$ values. The
$m_T$ spectra measured in $Au+Au$ are shown in Fig. 1(b)
and 1(c) for $e^{+}e^{-}$ and $K^{+}K^{-}$, respectively with 
a fit to an exponential function to extract the rapidity density
$dN/dy$ and the temperature T. The temperatures extracted in the two decay
channels are in good agreement with each other as seen in
Fig. 1(d), whereas an increase of the normalized yield from
$d+Au$ collisions to peripheral $Au+Au$ collisions can be seen for
both $e^+e^-$ and $K^+K^-$ in Fig. 1(e)~\cite{kozlov}. The
dielectron yield in the $Au+Au$ channel seems to be higher compared to
the dikaon channel, but the large systematic and statistical
uncertainties in the $e^+e^-$ measurement prevent us from making a
definitive statement. The measurements are expected to improve in
future with the newly installed Hadron Blind Detector~\cite{hbd}. The
$p_{T}$ spectra of $\omega$ meson measured using
$e^{+}e^{-},~\pi^{0}\gamma$ and $~\pi^{0}\pi^+\pi^-$ are shown in 
Fig. 1(f) and 1(g). The spectra show good agreement among the
different decay channels in all the three systems $p+p$, $d+Au$ and
$Au+Au$. The solid curve in Fig. 1(g) is the $p_{T}$ parameterization of
the $\pi^{0}$ yield measured in $p+p$ collisions and the dashed curves
are the $p+p$ fits scaled by the corresponding number of binary
collisions. The curves describe pretty well the measured spectra in
$d+Au$ and peripheral $Au+Au$ collisions, whereas for central $Au+Au$
collisions, the measured points lie below the curve, indicating some
suppression. 
%===========================================================================================================

\begin{figure}[!ht]
  \vspace{-3mm}
  \begin{center}
    \includegraphics[height=1.9in,width=2.5in]{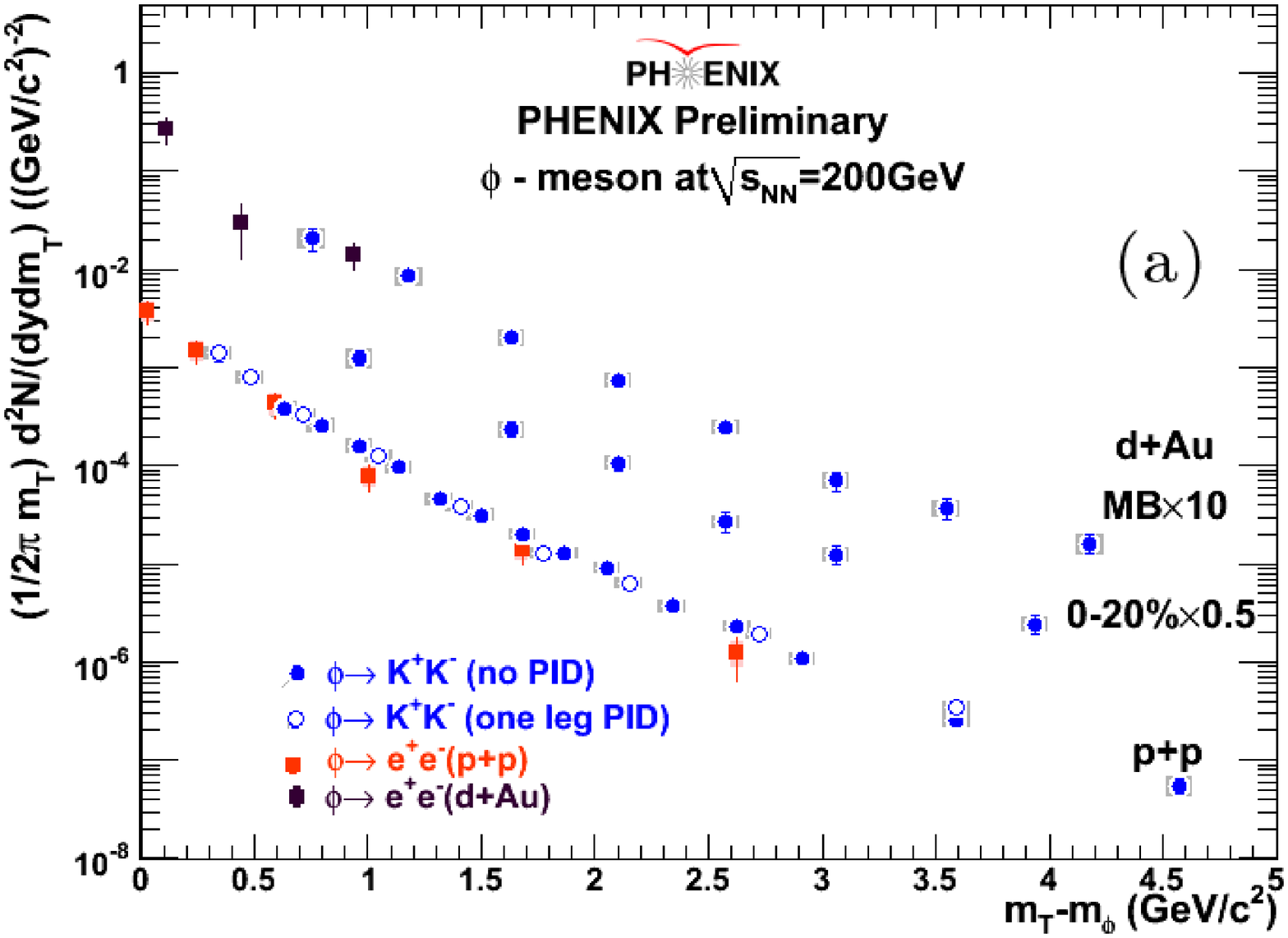}
    \includegraphics[height=1.9in,width=2.5in]{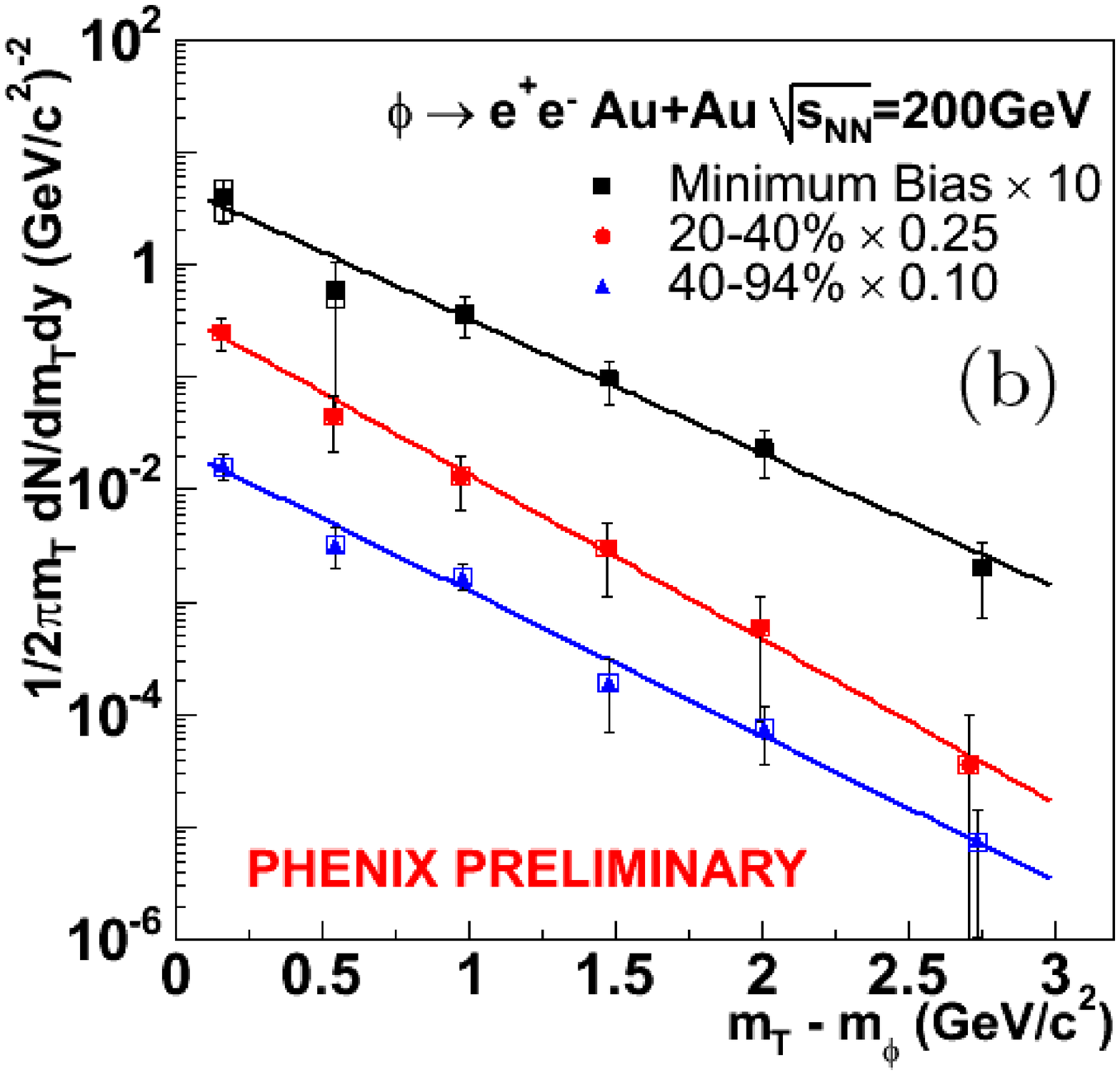}\\
    \includegraphics[height=1.9in,width=2.5in]{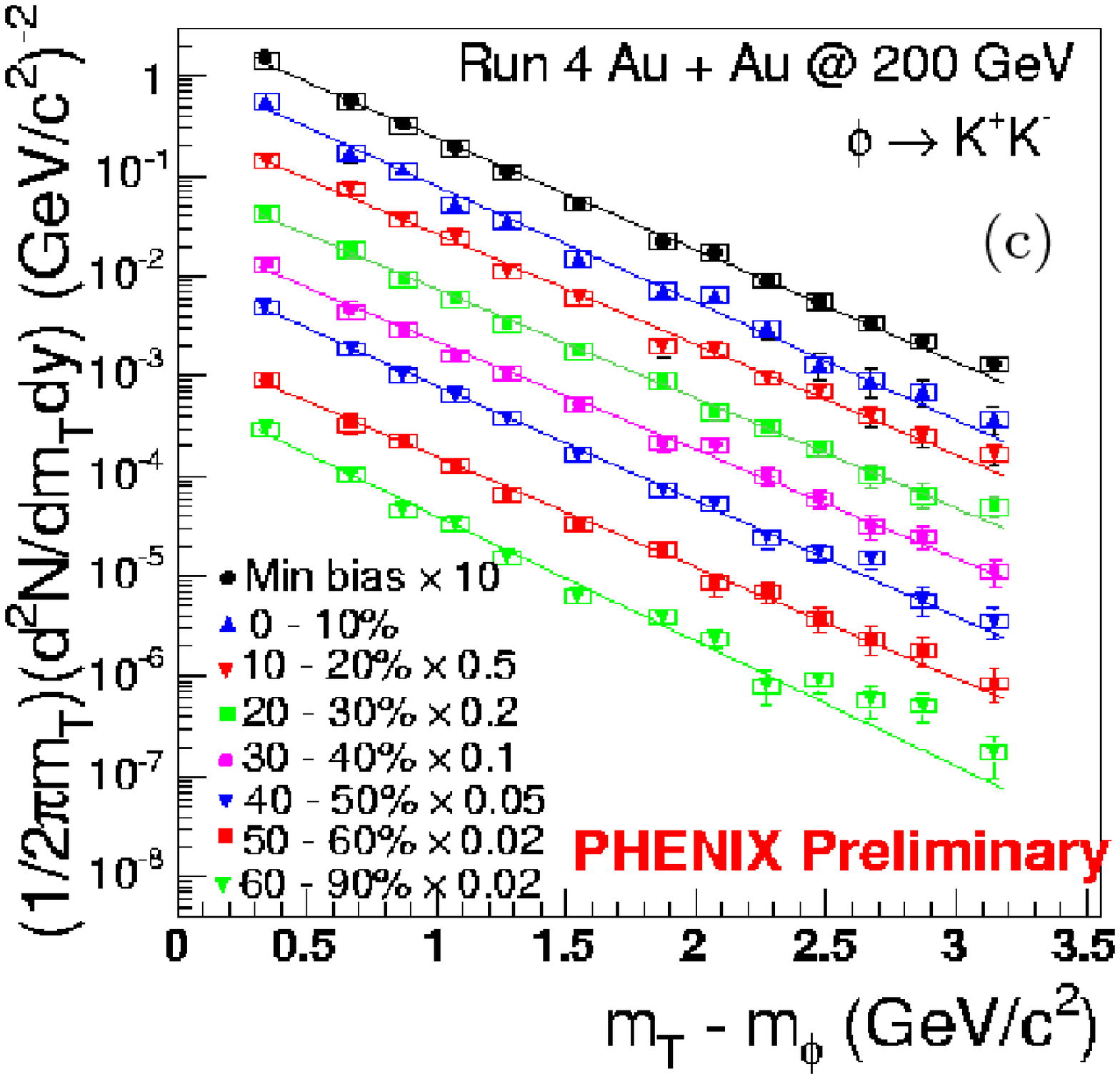}
    \includegraphics[height=1.95in, width=2.5in]{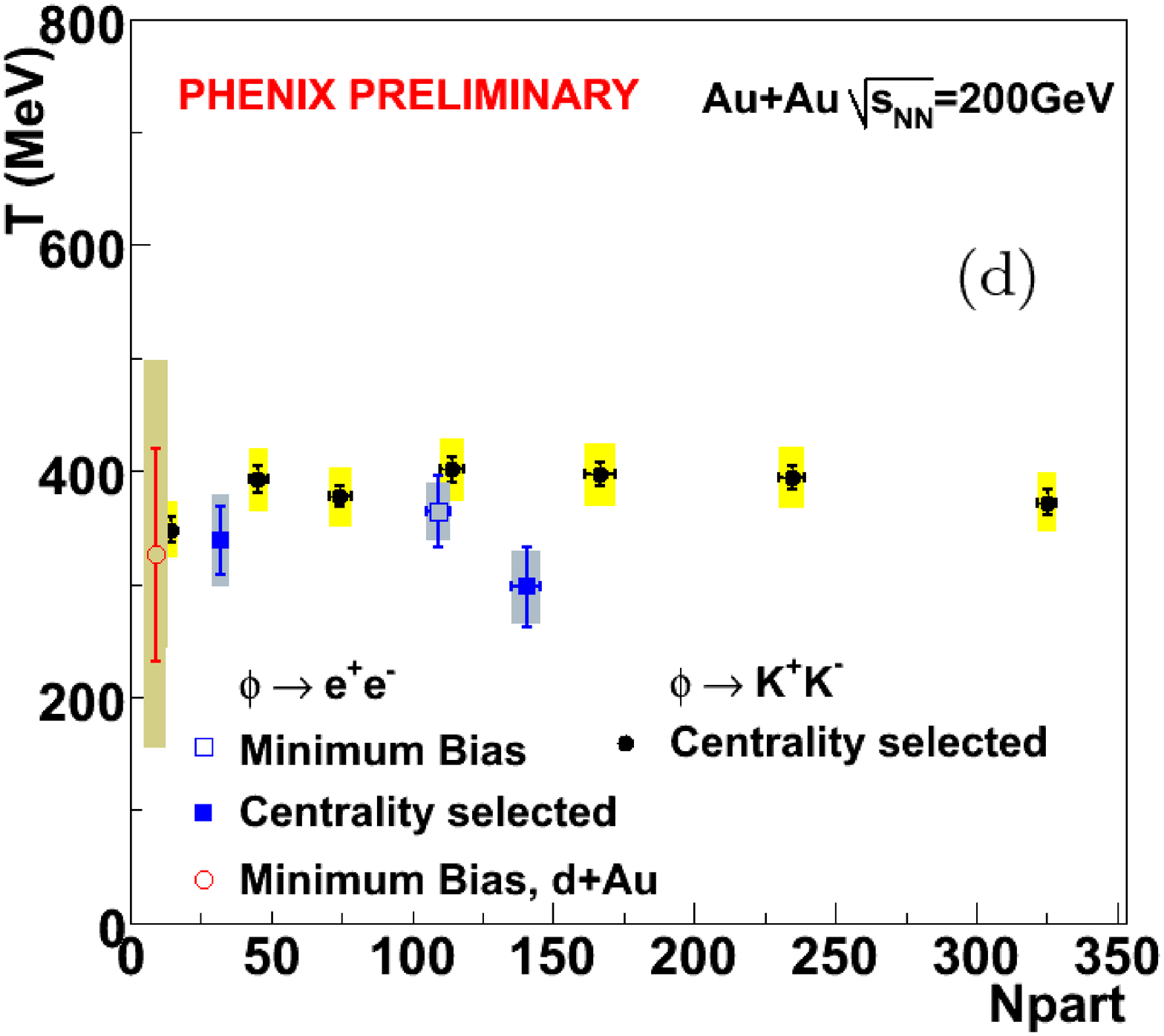}\\
    \includegraphics[height=1.9in, width=2.5in]{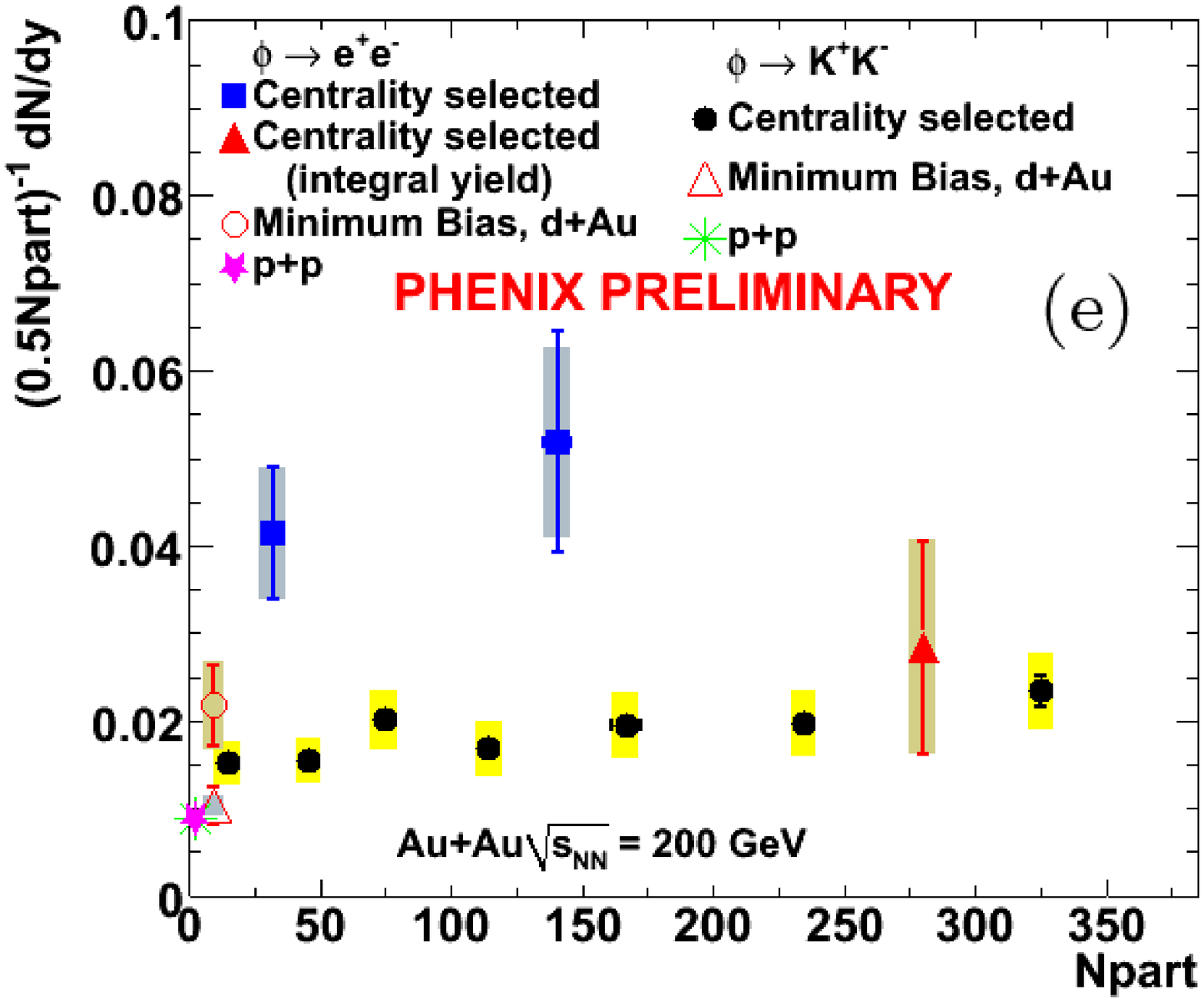}
    \includegraphics[height=1.9in,width=2.5in]{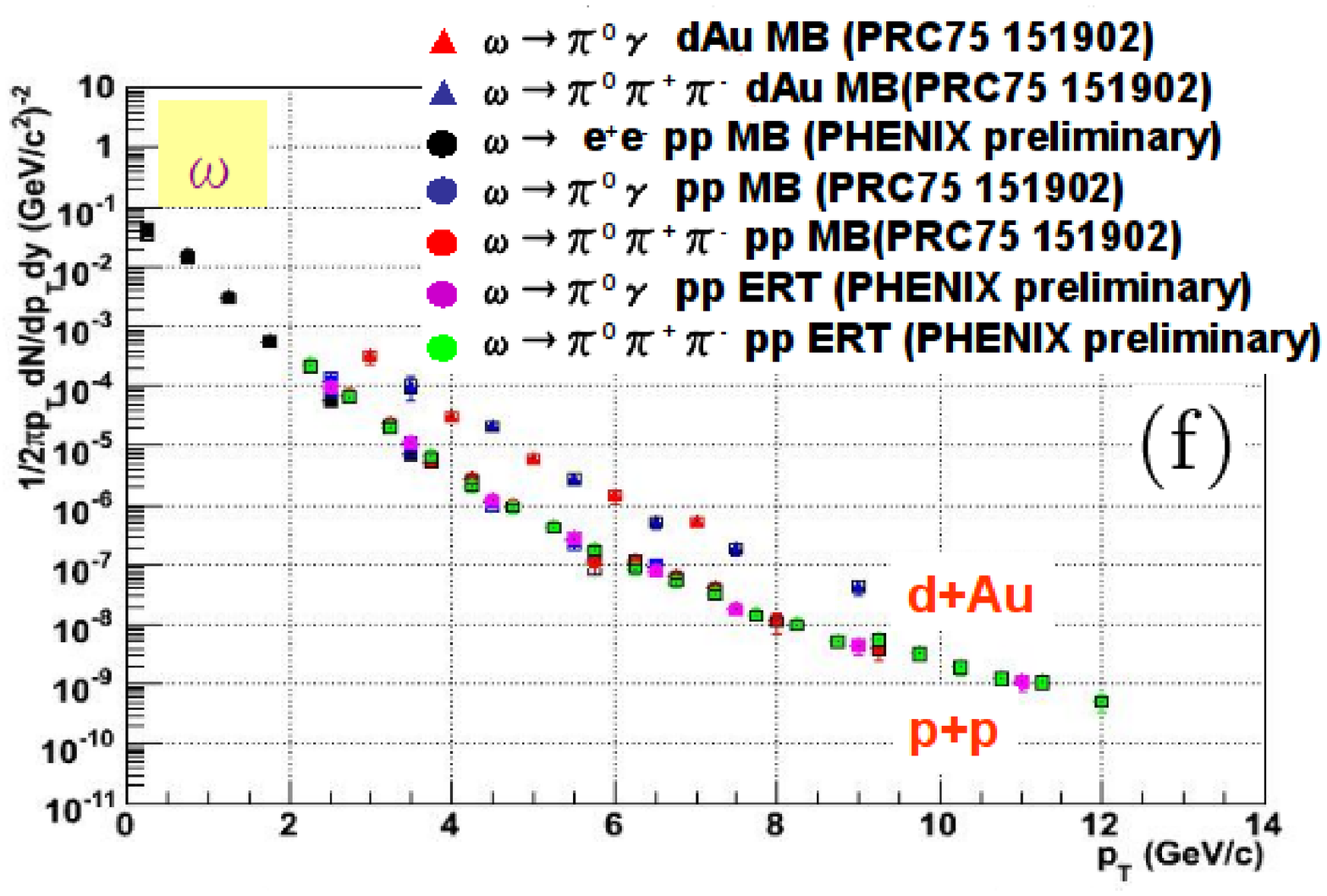}\\
    \includegraphics[height=1.9in,width=2.5in]{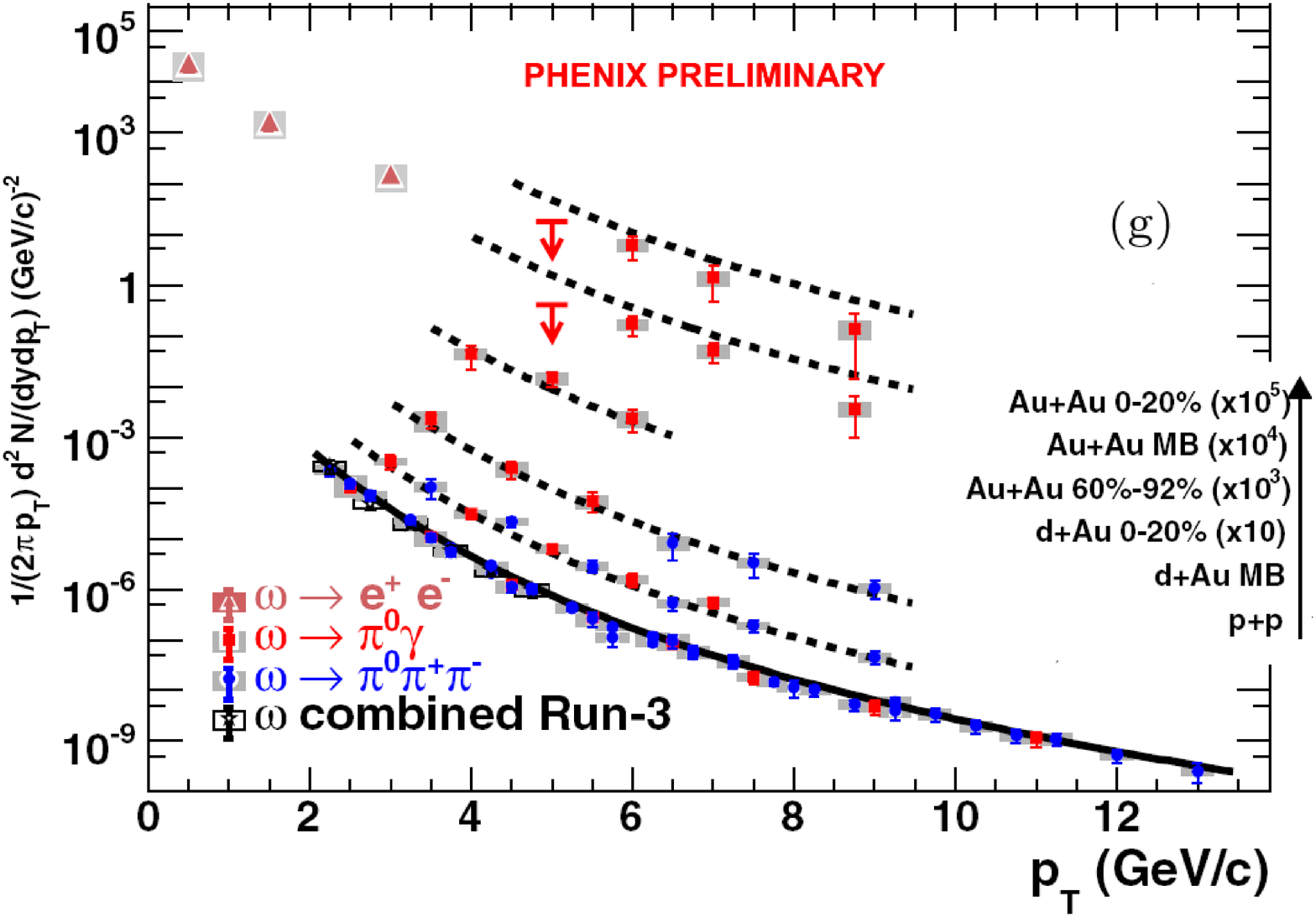}
  \end{center}
  \vspace{-5mm}
  \caption {$m_{T}$ spectra for $\phi$ in (a) $p+p$ and $d+Au$
    collisions via $e^+e^-$ and $~K^+K^-$, (b) $Au+Au$ ($e^+e^-$) and
    (c) $Au+Au$ ($K^+K^-$) (c). (d) and (e) show T and dN/dy for the two
    decay channels vs multiplicity. $p_{T}$ spectra of $\omega$ are
    shown in (f) for $p+p$ and $d+Au$ and in (g) for $Au+Au$ collisions.}
   \label{fig_all}
 \end{figure}
%===========================================================================================================

\end{document}